\documentclass[english,keywords,aps,twocolumn]{revtex4-1}
\usepackage{babel}
\usepackage{amssymb}
\usepackage{graphicx}
\usepackage{color}
\usepackage{bm}
\usepackage{longtable}
\usepackage{amsmath} 
\usepackage{amsfonts}
\usepackage{amssymb}
\usepackage{amsmath}%
\usepackage{amsfonts}%
\usepackage{amssymb}%
\usepackage[cal=cm]{mathalfa}
\usepackage{siunitx}
\usepackage{graphicx}
\usepackage{braket}
\usepackage{sidecap}
\usepackage{color}
\usepackage{bbm}
\usepackage{nicefrac}
\usepackage{float}
\usepackage[dvipsnames]{xcolor}
\usepackage[utf8]{inputenc}
\usepackage{xcolor, soul} 
\usepackage{sidecap}
\usepackage{relsize}
\usepackage{physics}
\usepackage{amsmath,amssymb,amsthm,mathrsfs,amsfonts,dsfont} 
\usepackage{braket}
\usepackage{bbold}
\usepackage{fancyhdr}
\usepackage{soul}

\newcommand{\im}{\text{i}}
\begin{document}

\title{Quantifying the presence of a neutron in the paths of an interferometer}

\author{Hartmut Lemmel$^{1,2}$}
\email{hartmut.lemmel@tuwien.ac.at}
\author{Niels Geerits$^{1}$}
\author{Armin Danner$^{1}$}
\author{Yuji Hasegawa$^{1}$}
\author{Holger F. Hofmann$^{3}$}
\author{Stephan Sponar$^{1}$}
\email{stephan.sponar@tuwien.ac.at}

\affiliation{
$^1$Atominstitut, TU Wien, Stadionallee 2, 1020 Vienna, Austria \\
$^2$Institut Laue-Langevin, 38000, Grenoble, France\\
$^3$Graduate School of Advanced Science and Engineering, Hiroshima University, Higashi Hiroshima 739-8530, Japan
}

\begin{abstract}
It is commonly assumed that no accurate experimental information can be obtained on the path taken by a particle when quantum interference between the paths is observed. However, recent progress in the measurement and control of quantum systems may provide the missing information by circumventing the conventional uncertainty limits. Here, we experimentally investigate the possibility that an individual neutron moving through a two-path interferometer may actually be physically distributed between the two paths. For this purpose, it is important to distinguish between the probability of finding the complete particle in one of the paths and the distribution of an individual particle over both paths.
We accomplish this distinction by applying a magnetic field in only one of the paths and observing the exact value of its effect on the neutron spin in the two output ports of the interferometer. The results show that individual particles experience a specific fraction of the magnetic field applied in one of the paths, indicating that a fraction or even a multiple of the particle was present in the path before the interference of the two paths was registered.
The obtained path presence equals the weak value of the path projector and is not a statistical average but applies to every individual neutron, verified by the recently introduced method of feedback compensation.

\end{abstract}

\maketitle

\section{Introduction}

Young's single-particle double-slit experiment is at the heart of wave-particle duality  \cite{deBroglie,Heisenberg27}, a particular form of complementarity \cite{Bohr28}, and one of the most fundamental concepts of quantum mechanics. The incompatibility between the wave-like behaviour of quantum objects and the \emph{classical} concept of a particle manifests in the observed aspect (wave or particle) and depends on the experimental context. When a position measurement to determine which slit a quantum particle traverses (particle-like property) is performed, in turn the interference pattern (wave-like property) vanishes; the more \emph{which-way} information is extracted, the lower is the visibility of the interference fringes and vice versa. This particular behavior is quantified in the Englert-Greenberger relation \cite{Englert96}.

Einstein ever since questioned this impossibility of determining the path taken by an individual particle in a double-slit interference experiment \cite{EinsteinBook27}. In his proposed scheme, which-way information is gained by measuring the recoil seemingly without destroying the interference pattern. However, as pointed out by Bohr, Einstein's proposal was in conflict with the principles of quantum mechanics. Nevertheless, this was the very first version of a \emph{welcher-weg} or \emph{which-way} thought experiment. In the following years different variants of such a \emph{which-way} experiment were considered, demonstrating the mutual exclusivity of which-way information and interference. 

A versions of Einstein's which-way thought experiment, that is compatible with the laws of quantum mechanics, was developed by Wootters and Zurek in 1979 \cite{Wootters79}.
This work paved the way for so called \emph{delayed-choice experiments}, first introduced by Wheeler \cite{WheelerBook}, where both scenarios (which-way information or interference) are still possible after a photon has passed the first plate of a Mach-Zehnder interferometer by removing or inserting the last interferometer plate.

A recently widely discussed which-way experiment was proposed by Vaidman in \cite{Vaidman13}, where which-path information is extracted from faint traces (with minimal perturbations) left along the beam path taken by the particle. Vaidman's so called \emph{past of a quantum particle} experiment was realized using photons \cite{Vaidman13} and later neutrons \cite{Geppert18}. A recent which-way experiment, applying a slightly different approach to the which-way problem, is reported in \cite{Fenghua20}. 

Except for Vaidman's past of a quantum particle experiment, all of these approaches evaluate the which path information without any reference to the initial quantum state, neglecting any possible correlations between the which-way information and the interference effects that might be described by the initial state.  This assumption is intuitively justified by the symmetry of conventional interference experiments, where both paths contribute equally to the interferences observed. However, this is not always the case. If there is an imbalance between the paths in the initial state, this imbalance includes correlations between which-way information and the outcome of an interference experiment, so that the individual outcomes of the interference measurements provide some non-trivial which-way information in addition to the observation of interference. This possibility of using correlations of the input state to evaluate physical properties based on arbitrary measurement outcomes is expressed by the state dependent uncertainties introduced by Ozawa in 2003 \cite{Ozawa03}. As explained by Hall, Ozawa's definition of measurement errors indicates that the \emph{best estimate} of a physical property for any measurement is given by the real part of the weak value defined by the initial state and the measurement outcome \cite{Hall04}. When this theory is applied to the which-way problem, it is possible to evaluate the presence of a particle in the two paths based on the outcome of an interference measurement. Somewhat surprisingly, the theory attributes an uncertainty of zero to optimal estimates of the path presence which can be a fraction of one or even larger than one when the initial state is biased in favour of the path in which the presence is evaluated.

Originally it was thought that the Ozawa-Hall uncertainties have no experimentally observable consequences \cite{Werner04}. Previous experiments reconstructed the uncertainties based on statistical assumptions that were motivated by a theoretical analysis of the formalism, either using a tomographic reconstruction, i.e., three-state-method \cite{Ozawa04,Erhart12,Edamatsu13,Sulyok13,Ringbauer14,Sulyok17}, or weak measurements \cite{Lund10,Steinberg12}. However, it has recently been shown in \cite{Hofmann21} that Ozawa-Hall uncertainties can be directly observed as the uncertainty in the rotation of a probe qubit when the method of feedback compensation is used, cf. next Section. We present an implementation of this method and
the results show that the weak values of the particle presence in the path associated with the output ports of the interferometer have extremely low errors and therefore apply with precision to every particle that exits the interferometer through that port. The evaluation of errors thus confirms that a single particle can be physically distributed between the paths in the context of an interference measurement, demonstrating that the path of a particle strongly depends on the measurement context established in the final measurement performed in the output ports.

The remaining paper is organized as follows: In Section II we introduce the theory of feedback compensation applied to a which-way measurement of a Mach-Zehnder interferometer. In Section III we compare the path presences obtained in the interference experiment with the ones of a conventional which-way measurement.
In Section IV we present the results of our neutron interferometric experiment in the two different experimental contexts. In Section IV we discuss the obtained results and their connection to the theoretical framework of quantum errors introduced by Ozawa.

\section{Theory and Setup} \label{sec:theory}

\begin{figure}
{\includegraphics[width=0.45\textwidth]{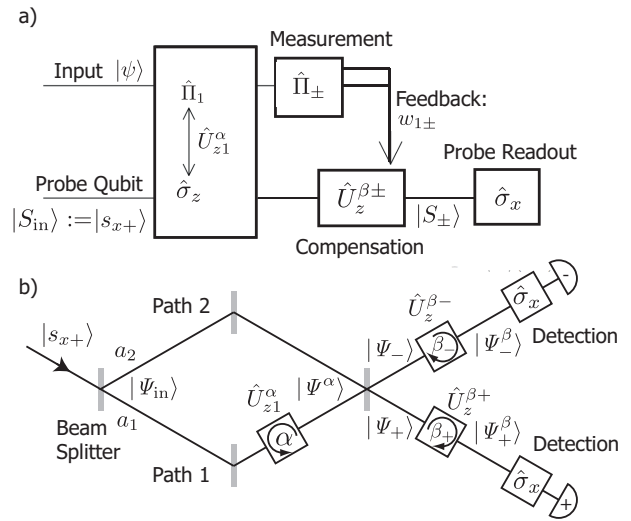}}
\caption{(a) Scheme of \emph{feedback compensation} from \cite{Hofmann21} as applied to a Mach-Zehnder interferometer (b).  After a coupling $\hat U^\alpha_{z1}$ between object (interferometer paths) and probe system (spin), a compensation $\hat U_z^{\beta\pm}$ dependent on the output channel is applied, maintaining the original state $\langle\hat\sigma_x\rangle=1$ of the probe qubit.  }\label{fig:schemeFeedback}
\end{figure}

The feedback compensation scheme \cite{Hofmann21}, illustrated in Fig.\,\ref{fig:schemeFeedback} (a) proposes to weakly couple a probe qubit to the observable of interest. Then the probe qubit carries information about the observable, and particularly precise information can be gained if the probe qubit is finally changed back to its original state by applying a compensation operation. The compensation depends on the measurement outcome of the observable and is therefore called feedback compensation. We explain the scheme in detail on the basis of our particular setup. The setup is schematically shown in Fig.\,\ref{fig:schemeFeedback} (b) and the indicated quantum states are described in the following.

We send neutrons into the interferometer where we assume for universality and for illustrative reasons an asymmetric beam splitter. The neutron in the interferometer is then described by the state
\begin{alignat}{1}
  |\psi\rangle &= 
    a_1 |1\rangle + a_2 |2\rangle
\end{alignat}
where $|1\rangle$ and $|2\rangle$ denote the eigenstates of path 1 and 2 respectively. $a_1$ and $a_2$ are normalized real amplitudes, $a_1^2 + a_2^2=1$. The two exit ports of the interferometer are described by the states $|+\rangle$ and $|-\rangle$ respectively, which also take the phase $\chi$ between the two paths into account.
\begin{alignat}{1}
  |\pm\rangle &= \frac 1 {\sqrt 2}\bigl(\,|1\rangle \pm \exp(\im\chi)\, |2\rangle\bigr)
\end{alignat}
We want to gain information about which path the neutron takes and use the neutron spin as probe qubit. We prepare it initially in the $|s_{x+}\rangle$ state and the total initial state reads
\begin{subequations}
\begin{alignat}{1}\label{eq:initialState}
  |\mathit\Psi_{\rm{in}}\rangle &= |\psi\rangle     |s_{x+}\rangle \\
  |s_{x+}\rangle &= \frac 1 {\sqrt 2} \left( |\uparrow\rangle + |\downarrow\rangle \right)
\end{alignat}
\end{subequations}
where $|\uparrow\rangle$ and $|\downarrow\rangle$ denote the spin eigenstates in $z$ direction. The $z$ basis is the natural choice of spin representation since we use an external guide field in $z$ direction. 

We couple spin and path degree of freedom by weakly rotating the spin in path 1. We rotate it by a small angle $\alpha$ about the $z$ axis. The rotation is expressed by the operator $\hat U_z^{\alpha}$ acting only in path 1 or, equivalently, by the operator $\hat U_{z1}^{\alpha}$ acting on the total state of both paths and spin.
\begin{subequations}
\begin{alignat}{1}
  \hat U_{z1}^\alpha &= \exp \left(-\frac \im 2 \alpha \, \hat \sigma_z \,  \hat \Pi_1 \right) = 
 \hat \Pi_1  \hat U_{z}^\alpha + \hat \Pi_2  \\
  \hat U_{z}^\alpha &=  \exp \left(-\frac \im 2 \alpha \, \hat \sigma_z\right) = \hat{\mathbb 1} \cos \frac \alpha 2 - \im \hat \sigma_z \sin \frac \alpha 2
\end{alignat}
\end{subequations}
where $ \hat \Pi_1$ and $ \hat \Pi_2$ denote the path projection operators of path 1 and 2 respectively. 
The identities between the exponentials and the right hand terms can be proven by series expansion of the exponential and the fact that $\hat\sigma_z^2=\hat{\mathbb 1}$ and $\hat \Pi_{1,2}^2 = \hat \Pi_{1,2}$. The state after the spin rotation reads
\begin{alignat}{1}
  |\mathit\Psi^\alpha\rangle &=  \hat U_{z1}^\alpha |\mathit\Psi_{\text{in}}\rangle = \hat U_{z1}^\alpha   |\psi\rangle     |s_{x+}\rangle.
\end{alignat}
The states of the two exit beams of the interferometer are given by the projection onto the exit states $|\pm\rangle$ respectively.
\begin{alignat}{1}
  |\mathit\Psi_\pm\rangle &=|\pm\rangle  \langle\pm |\mathit\Psi^\alpha\rangle =  \langle\pm | \hat U_{z1}^\alpha  |\psi\rangle \; |\pm\rangle    |s_{x+}\rangle
\end{alignat}
In the exit beams we rotate the spin back by the angle $\beta$. This is the compensation. It acts likewise on the components coming from both paths 1 and 2, but, in general, different $\beta$ angles will be chosen for the exit beams $|+\rangle$ and $|-\rangle$ because the optimal compensation depends on the measurement outcome, i.e. in which of the two exit beams the neutron is eventually detected. The compensated state then reads
\begin{alignat}{1}
  |\mathit\Psi_\pm^\beta\rangle &= \hat U_z^{-\beta}   |\mathit\Psi_\pm\rangle = \langle\pm | \hat U_z^{-\beta} \hat U_{z1}^\alpha  |\psi\rangle \;  |\pm\rangle    |s_{x+}\rangle.
\end{alignat}
By writing the operators as 
\begin{alignat}{1}
  \hat U_z^{-\beta} \hat U_{z1}^\alpha &=  
  \hat\Pi_1 \hat U_z^{\alpha-\beta } +
  \hat\Pi_{2} \hat U_z^{-\beta}
\end{alignat}
we clearly see that the spin of the path-1 component is rotated by $\alpha-\beta$ while the spin of the path-2 component is rotated by $-\beta$. The final components observed in the exit beams $|+\rangle$ and $|-\rangle$ can be written as
\begin{alignat}{1}
  |\mathit\Psi_\pm^\beta\rangle =  \langle\pm | \psi\rangle 
  \left( \omega_{1\pm} \hat U_z^{\alpha-\beta } + \omega_{2\pm}  \hat U_z^{-\beta}\right) \; |\pm\rangle | s_{x+}\rangle 
\end{alignat}
where $\omega_{1\pm}$ and $\omega_{2\pm}$ denote the weak values \cite{Aharonov88,Sponar15} of the path projection operators respectively. 
\begin{alignat}{1}
  \omega_{1\pm} &=  \frac{\langle\pm | \hat\Pi_1 | \psi\rangle }{\langle\pm | \psi\rangle } = \frac 1 {1 \pm  \frac {a_2}{a_1} \exp(\im\chi)} \nonumber \\
  \omega_{2\pm} &= \frac{\langle\pm | \hat\Pi_2 | \psi\rangle }{\langle\pm | \psi\rangle } = \frac 1 {1 \pm \frac {a_1}{a_2}\exp(-\im\chi)}  = 1 - \omega_{1\pm} \label{eq:omega}
\end{alignat}
Remark: In case of a symmetric beam splitter ($a_1=a_2$) and $\chi=0$ all neutrons reach the $|+\rangle$ state. The amplitude $\langle - | \psi\rangle$ of the $|-\rangle$ state vanishes and the weak values $\omega_{1,2-}$ diverge. 

To read out the probe qubit we analyze the spin 
in the two output ports. We factorize the final state into the path dependent part $\langle\pm | \psi\rangle |\pm\rangle$ and the spin dependent part $|S_\pm\rangle$.
\begin{subequations}
\begin{alignat}{1}
  |\mathit\Psi_\pm^\beta\rangle &= \langle\pm | \psi\rangle \, |\pm\rangle |S_\pm\rangle \\
  |S_\pm\rangle&= \left(\omega_{1\pm} \hat U_z^{\alpha-\beta} +  \omega_{2\pm} \hat U_z^{-\beta} \right)| s_{x+}\rangle
\end{alignat}
\end{subequations}
We calculate the amplitudes of the $\hat\sigma_x$ eigenstates of the spin qubit.
\begin{alignat}{1}
  s_{x+}&= \langle s_{x+}|S_\pm\rangle
 =\omega_{1\pm} \cos \frac {\alpha-\beta} 2 + \omega_{2\pm} \cos \frac {\beta} 2\label{eq:sxplus}\nonumber\\
 s_{x-}&=  \langle s_{x-}|S_\pm\rangle
   = \im \left(-\omega_{1\pm} \sin \frac {\alpha-\beta} 2 + \omega_{2\pm} \sin \frac {\beta} 2\right) 
\end{alignat}
With the substitution
\begin{alignat}{1}
  A_\pm\, \sin\frac{\beta_{0\pm}}2 &:=  \omega_{1\pm} \sin\frac\alpha 2 \nonumber\\
  A_\pm \,\cos\frac{\beta_{0\pm}}2 &:= \omega_{1\pm} \cos\frac\alpha 2 + \omega_{2\pm}
\end{alignat}
we see that the spin amplitudes $s_\pm$ oscillate as function of the compensation angle $\beta$ with a certain phase $\beta_{0\pm}$ and with an oscillation amplitude $A_\pm$.
\begin{alignat}{1}
  s_{x+} &= A_\pm \, \cos \frac{\beta - \beta_{0\pm}}2 \nonumber\\
  s_{x-} &= \im A_\pm \, \sin \frac{\beta - \beta_{0\pm}}2 \label{eq:sxbeta}
\end{alignat}
\begin{alignat}{1}
  \beta_{0\pm} &= 2 \arctan \frac{\sin \frac \alpha 2}{\frac {\omega_{2\pm}}{\omega_{1\pm}} + \cos \frac \alpha 2 } \label{eq:beta0pm}\\
  A_\pm^2 &= 1-4 \omega_{1\pm} \omega_{2\pm} \sin^2 \frac\alpha 4 \label{eq:Apm}
\end{alignat}
Eqs. (\ref{eq:sxbeta}) show that perfect compensation is obtained if $\beta = \beta_{0\pm}$. Then $s_{x-}$ vanishes and the final state equals the initial state $|s_{x+}\rangle$. The consequences are discussed further below.

A series expansion of $\beta_{0\pm}$ by $\alpha$ shows that $\beta_{0\pm}$ is to first order determined by the weak value $\omega_{1\pm}$.
\begin{alignat}{1}
 \beta_{0\pm} &=  \omega_{1\pm} \alpha 
 + \mathcal{O}( \alpha^3)
 \label{eq:beta0pmSeries}\\
 A_\pm &= 1-\frac 1 2 \omega_{1\pm} \omega_{2\pm} \left(\frac\alpha 2\right)^2 + \mathcal{O}( \alpha^4)  \label{eq:ApmSeries}
\end{alignat}
In other words, in the limit of weak coupling (small $\alpha$) the weak value of the path projection operator is given by the ratio of the spin rotation angle $\alpha$ and the optimal compensation $\beta_{0\pm}$.
\begin{alignat}{1}
 \omega_{1\pm} &= \beta_{0\pm} /  \alpha
\end{alignat}
This can also be concluded from a first order series expansion of the operators.
\begin{alignat}{1}
  \hat U_z^\beta \hat U_{z1}^\alpha &=  \exp \left(-\frac {\im\alpha} 2 \left( \hat \Pi_1 - \frac \beta \alpha  \right) \, \hat \sigma_z \right)\nonumber  \\ &\approx
\hat{\mathbb 1} - \frac {\im \alpha} 2 \hat \sigma_z \left( \hat \Pi_1 - \frac \beta \alpha \right)
\end{alignat}

The weak values $\omega_{1\pm}$ can eventually be determined by analyzing the spin expectation values $\langle \hat \sigma_x \rangle$, $\langle \hat \sigma_y \rangle$ and $\langle \hat \sigma_z\rangle$. (See App. \ref{app:expect} for a detailed derivation.)
\begin{subequations}
\begin{alignat}{1}
  \langle \hat \sigma_x \rangle &= 
  \frac {\cos (\Re\beta_{0\pm}-\beta)}{\cosh \Im \beta_{0\pm}}  \label{eq:sigmax}\\
  \langle \hat \sigma_y \rangle &= 
  \frac {\sin (\Re\beta_{0\pm}-\beta)}{\cosh \Im \beta_{0\pm}}  \\
  \langle \hat \sigma_z \rangle &=
  \tanh \Im \beta_{0\pm}
\end{alignat}
\end{subequations}
A second order series expansion by $\alpha$ gives 
\begin{subequations}
\begin{alignat}{1}
 \langle \hat \sigma_x \rangle &= 
  \cos (\beta - \alpha\Re\omega_{1\pm}) \, \left( 1 - \frac 1 2 (\alpha \Im \omega_{1\pm})^2 \right) \label{eq:sxexpectweak}\\
 \langle \hat \sigma_y \rangle &= 
  \sin (\beta - \alpha\Re\omega_{1\pm}) \, \left( 1 - \frac 1 2 (\alpha \Im \omega_{1\pm})^2 \right) \\
\langle \hat \sigma_z \rangle 
  &= \alpha\, \Im \omega_{1\pm}. 
\end{alignat}
\end{subequations}
In a conventional experiment, no compensation is applied ($\beta$=0) and the weak value is determined for example as $\omega_1 \approx \frac 1 \alpha (-\langle\sigma_y\rangle + \im \langle\sigma_z\rangle)$ \cite{Sponar15,Vallone16,Denkmayr17}. 
In order not to disturb the interference, the angle $\alpha$ is kept small, and the measurement is called a weak measurement. Due to the complementarity of which-way information and visibility of interference \cite{Englert96}, only little information can be gained per event and many events have to be measured. Since the spin state is close to $|s_{x+}\rangle$ the variances of $\langle\sigma_y\rangle$ and $\langle\sigma_z\rangle$ are close to their maximum (cf. App. \ref{app:variances}), and the obtained weak values are clearly a statistical average over the whole ensemble. They don't tell anything about an individual neutron. 

In the feedback compensation scheme, we don't measure individual spin components but determine the spin rotation relative to its initial state. This is done by applying the compensation, i.e. an estimated back rotation by $\beta$, and measuring the spin in the original direction $x$. The estimate is varied until $\langle\sigma_x\rangle$ reaches a maximum. In an ideal case $\max(\langle\sigma_x\rangle)=1$ and the variance of $\langle\sigma_x\rangle$ between events vanishes completely. Complementarity is still valid in the sense that many events are needed to determine the optimal compensation. But once the compensation is adjusted, every neutron verifies its correctness and the determined  weak value can be attributed to every individual neutron.

A general concept of measurement errors was introduced by Ozawa \cite{Ozawa03} and it was shown by Hall that these measurement errors correspond to the uncertainty of an estimate of a physical property based on the outcome of an arbitrary measurement \cite{Hall04}. This uncertainty is given by the statistical deviation between the operator of interest and the estimated value of that operator. In our case it reads
\begin{alignat}{1}
  \varepsilon^2(\hat \Pi_1) &= \langle\psi| 
\left[ \hat\Pi_1 \; - \; \sum_\pm \frac{\beta_\pm}\alpha |\pm\rangle\langle\pm|   \right]^2 
|\psi\rangle. \label{eq:epsilon}
\end{alignat}
As shown in \cite{Hofmann21}, this uncertainty describes the error of the feedback compensation for any choice of estimates $\beta_{\pm}/\alpha$. The experimentally observable result of this error is a reduction of the spin component $\langle\sigma_x\rangle$ from its original value of 1 to
\begin{alignat}{1}
 \langle\sigma_x\rangle=1 - \frac 1 2 \alpha^2 \varepsilon^2(\hat \Pi_1).\label{eq:maxsigmax}
\end{alignat}
In our experiment, the measurement basis $\{|+\rangle,|-\rangle\}$ is complete and orthogonal and the operator $\hat\Pi_1$ is self-adjoint. Then the uncertainty Eq. (\ref{eq:epsilon}) is determined only by the differences between the estimates and the weak values.
\begin{alignat}{1}
\varepsilon^2(\hat \Pi_1)
&= \sum_\pm \langle\psi| \left(\hat\Pi_1^\dagger - \frac{\beta_\pm}\alpha \right)|\pm\rangle\langle\pm |\left(\hat\Pi_1 - \frac{\beta_\pm}\alpha \right)|\psi\rangle\nonumber\\
  &= \sum_\pm p_\pm 
  \left|w_{1\pm} - \frac{\beta_\pm}\alpha \right|^2 \label{eq:uncertainty}
\end{alignat}
where $p_\pm$ denotes the statistical probability of finding the neutron in the final state $|\pm\rangle$ respectively.
\begin{alignat}{1}
p_\pm=|\langle \pm | \psi\rangle|^2=\frac 1 2 \pm a_1 a_2 \cos\chi 
\end{alignat}
This means, the uncertainty vanishes completely if the compensations $\beta_{\pm}/\alpha$ applied in the output ports $|+\rangle$ and $|-\rangle$ respectively equal the corresponding weak values $\omega_{1\pm}$. Then the compensations are no longer just estimates but precise measurements of the weak values. 

Since the compensation angles can only be real numbers, perfect compensation is only possible if the imaginary part of the weak values vanishes. This can be achieved by adjusting the phase $\chi$ between the two paths to zero, cf. Eq. (\ref{eq:omega}). In the following considerations and in the experiment we will focus on this case only.

\section{Path presence}

Let us assume that the real part of the weak values $\omega_{1}$ and $\omega_{2}$ of the path projectors describe physical reality, namely the presence of a neutron in the respective path. The effective spin rotation is then given by the angle $\alpha$ times the path presence in path 1, and it is no surprise that this rotation can be compensated by $\beta_{0\pm} = \alpha \, \omega_{1\pm}$. 

As shown before, the path presence it not a statistical average but can be attributed to every single detected neutron. We can exclude the possibility that some neutrons have taken only one path and other neutrons have taken only the other path and that they are distributed over the paths only in a statistical way. The probe qubit would carry on the noise of the path presence which is not the case. The path presence is precisely measured and it quantifies how each individual neutron was distributed between the paths.

The path presence should not be confused with the amplitudes $a_1$ and $a_2$ of the initial state. The latter determine the detection probabilities $p_{1,2}=|a_{1,2}|^2$ in case a neutron detector is placed directly in one or the other path. This would be a naive which-way measurement which destroys the interference. 
In contrast, the path presence proposed here can be precisely measured while full interference is maintained. Being a weak value, the path presence always links an initial state to a final state. This means, one still cannot predict the path a neutron will take, but once the neutron has been detected in one or the other exit beam, one can in retrospect infer its presence in path 1 and 2 respectively. 

While we cannot predict the path, we can calculate the statistical probability of detecting the neutron in path 1, which is given by the expectation value of the path projector $\hat\Pi_1$. In the context of the which-way measurement we measure in the eigenbasis of the operator, and obtain the expectation value by summing over the eigenvalues.
\begin{alignat}{1}
  \langle\psi |\hat\Pi_{1}|\psi\rangle &= |\langle\psi|1\rangle|^2 \cdot 1 +  |\langle\psi|2\rangle|^2 \cdot 0 = p_1
\end{alignat}
Since an expectation value can be calculated by any complete basis, we chose the $|\pm\rangle$ basis for the interference context. Now the weak values play the role of the eigenvalues, and the expectation value is given by the averaged weak value \cite{Hosoya10,Hall16}.
\begin{alignat}{1}
 p_1= \langle\psi |\hat\Pi_{1}|\psi\rangle &= |\langle\psi|+\rangle|^2 \cdot \omega_{1+} +  |\langle\psi|-\rangle|^2 \cdot \omega_{1-} \nonumber\\
  &= p_+ \omega_{1+}  + p_- \omega_{1-} 
  = \bar \omega_{1}  \label{eq:p1}
\end{alignat}
We see, that the detection probability $p_1$ can also be interpreted as an averaged weak value $\bar\omega_1$.

Also the variance can be calculated in either way. In the which-way context we get
\begin{alignat}{1}
  \mathit\Delta^2(p_{1}) &= p_1 \, (1 - p_{1})^2   + p_2 \,  (0 - p_{1})^2 = p_1 p_2 \label{eq:Deltap1}
\end{alignat}
and in the interference context
\begin{alignat}{1}
  \mathit\Delta^2(\bar \omega_{1}) &= p_+ (\omega_{1+} - \bar \omega_{1})^2   + p_-  (\omega_{1-}- \bar \omega_{1})^2
 = p_1 p_2. \label{eq:Deltaomega}
\end{alignat}
In short, the statistics of the input state is described equally by eigenvalues in the which-way context and by weak values in the interference context \cite{hofmann2020}. We can also define weak values in the which-way context $\omega_{pf}=\langle f |\hat\Pi_p|\psi\rangle/ \langle f | \psi \rangle$ where $p$ denotes the path of the weak value and $f$ the path where the neutron has been detected. However, these weak values coincide with the eigenvalues $\omega_{pf}=\delta_{pf}$. Weak values are the more general concept, respecting the actual outcomes of a measurement even if they are not represented by eigenstates. We take this as a strong hint that the weak values can really be interpreted as presence of the neutron in the respective path.

Table \ref{tab:pathPresences} shows a comparison of (a) the initial state, (b) the description in the interference context and (c) in the which-way context. As an example we have calculated the numbers for a $4:1$ beam splitter. In the interference context, the neutrons reach the $|+\rangle$ and $|-\rangle$ ports with the probabilities $p_+=0.9$ and $p_-=0.1$ respectively. A neutron ending in the $|+\rangle$ port has been to two thirds in path 1 and to one third in path 2. However, a neutron ending in the $|-\rangle$ port had a path-1 presence of $2$ and a path-2 presence of $-1$. 
These are so-called anomalous weak values which lie outside the eigenvalue spectrum of the path projection operator. The present experiment confirms that such anomalous weak values are necessary \cite{Hosoya10,Pusey14} to describe the path presence created by the input state Eq. (\ref{eq:p1}) and its fluctuations Eq. (\ref{eq:Deltaomega}) by the outcome probabilities of the interference experiment.

\begin{table}
\begin{tabular}{ l | c | c c }
 \multicolumn{2}{l|}{(a)}  & path 1 & path 2 \\ 
 \multicolumn{2}{l|}{   }  &        &        \\[-10pt]
\hline
 \multicolumn{2}{l|}{initial amplitudes} & $a_1=\frac 2 {\sqrt 5}$ & $a_2=\frac 1 {\sqrt 5}$\\
 \multicolumn{2}{l|}{initial probabilities} & $p_1=\frac 4 5 $ & $ p_2=\frac 1 5$ \vspace{16pt}\\
(b)& & presence & presence \\[-4pt]
& probability & in path 1 & in path 2 \\
  \hline
 final $|+\rangle$ & $p_+=\frac 9 {10}$ & $\omega_{1+}=\frac 2 3$ & $\omega_{2+}=\frac 1 3$ \\
 final $|-\rangle$  & $p_-=\frac 1 {10}$ & $\omega_{1-}=2             $ & $\omega_{2-}=-1$ \\
  average& & $\bar \omega_1 = \frac 4 5 $ & $ \bar \omega_2=\frac 1 5$\\
  variance& & $\mathit\Delta(\bar \omega_1) = \frac 2 5 $ & $ \mathit\Delta(\bar \omega_2)=\frac 2 5$ \vspace{16pt}\\
(c)& & presence & presence \\[-4pt]
& probability & in path 1 & in path 2 \\
  \hline
 final $|1\rangle$ & $p_1=\frac 4 5$ & $\omega_{11}=1$ & $\omega_{21}=0$\\
 final $|2\rangle$ & $p_2=\frac 1 5$ & $\omega_{12}=0$ & $\omega_{22}=1$\\
  average& & $\bar \omega_1 = \frac 4 5 $ & $ \bar \omega_2=\frac 1 5$\\
  variance& & $\mathit\Delta(\bar \omega_1) = \frac 2 5 $ & $ \mathit\Delta(\bar \omega_2)=\frac 2 5$\vspace{5pt}
\end{tabular}
\caption{Path presences calculated for a $4:1$ beam splitter. (a) initial preparation, (b) path presences in the interference context, (c) path presences in a which-way measurement.	}\label{tab:pathPresences}
\end{table}

\begin{figure*}
\centering
\includegraphics[width=0.98\textwidth]{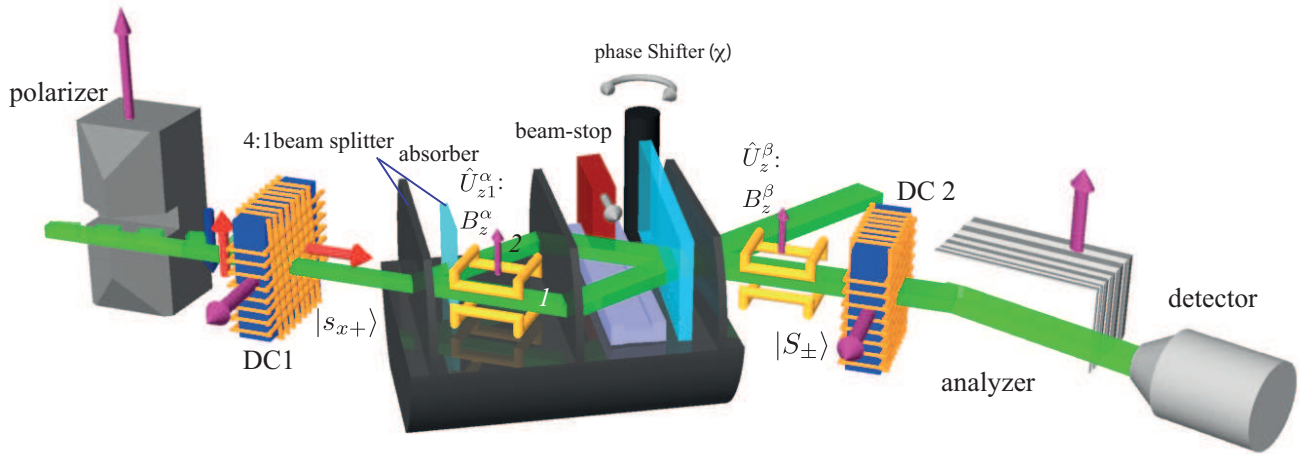}
\caption{Polarized monochromatic neutrons enter the interferometer and are split into paths  $\vert  1 \rangle$ and $\vert 2 \rangle$ at the first interferometer plate at a ratio of 1:4. Before the interferometer, the probe qubit is prepared in the initial spin state $\vert S_{\rm{in}}\rangle=\vert s_{x+}\rangle$ by a $\pi/2$ direct-current spin rotator  (DC\,1).  In path $1$ the spin is rotated by an angle $\alpha$. Beam stop and phase shifter are required for \emph{which-path} or \emph{interference} measurements, respectively. Behind the interferometer, the ``compensation'' is applied, that is a spin rotation by angle $\beta$, dependent of the respective measurement context. The spin is analyzed in $\pm x$ direction by the combination of a $\pi/2$ direct-current spin rotator (DC\,2) and the magnetic supermirror. The neutrons are counted in a $^3$He detector.}\label{fig:setup}
\end{figure*}

The path presences can be used to correctly calculate e.g. the average spin rotation angle in either context.
\begin{alignat}{6}
  \bar \alpha &= p_+ &&\omega_{1+} &&\alpha  &&+ p_- &&\omega_{1-} &&\alpha \nonumber  \\
  &= p_1 &&\omega_{11} &&\alpha  &&+ p_2 &&\omega_{12} &&\alpha   = \bar \omega_1 \alpha
\end{alignat}

Also the expectation value $\langle \hat \sigma_{x}\rangle$  measured in our experiment depends on the final state. In case we could not distinguish between the two exit beams, we could apply only a common compensation angle and measure only the averaged expectation value, denoted by $\overline{\langle\hat\sigma_x\rangle}_{\pm}$ in the interference context and $\overline{\langle\hat\sigma_x\rangle}_{12}$ in the which-way context. 
\begin{subequations}
\begin{alignat}{1}
  \overline{\langle\hat\sigma_x\rangle}_{\pm} &=  p_+ \langle\hat\sigma_{x}\rangle_+  + p_-  \langle \hat\sigma_{x}\rangle_-\\
 \overline{\langle\hat\sigma_x\rangle}_{12} &=  p_1 \langle\hat\sigma_{x}\rangle_1  + p_2  \langle \hat\sigma_{x}\rangle_2 
\end{alignat}
\end{subequations}
The detailed expressions are given in App. \ref{app:averaged}. In the limit of small $\alpha$ both expectation values converge to
\begin{alignat}{1}
 \overline{\langle\hat\sigma_x\rangle}_{\pm} = 
 \overline{\langle\hat\sigma_x\rangle}_{12} &= 
  \left(1 - \frac 1 2 p_1 p_2 \alpha^2\right)  \cos(\beta - \bar\alpha) \label{eq:meansigmax}
\end{alignat}
and have exactly the form predicted by Eq. (\ref{eq:maxsigmax}). 
Comparing both equations  we  identify a measurement error of $\varepsilon^2(\hat \Pi_1)  = p_1 p_2 =\mathit\Delta^2(p_{1})$. This means, by not distinguishing between the measurement outcomes we end up with a measurement error equal to the original path uncertainty Eq. (\ref{eq:Deltap1}).

\section{Experiment}

In the experiment we verified the path-1 presences listed in Table \ref{tab:pathPresences} and the reduction of $\max(\langle\sigma_x\rangle)$ expressed in Eq. (\ref{eq:meansigmax}) in case the exit beams are not distinguished. 

The experimental realization of the operations described in the theory (Sec. \ref{sec:theory}) are mostly straight forward and depicted in Fig. \ref{fig:setup}. The neutrons are polarized by a magnetic prism which deflects the spin-down neutrons out of the Bragg acceptance angle of the interferometer crystal. The spin rotator DC1 rotates the remaining spin-up neutrons by $\pi/2$ into the initial $|s_{x+}\rangle$ state. The asymmetry of the beam splitter is realized by an absorber in path 2. The spin rotations $\alpha$ and $\beta$ are realized by small Helmholz coils which modify the external overall $B_z$ guide field (not depicted in Fig.\,\ref{fig:setup}) such that the spin precession in the $x$-$y$-plane changes. The precession angle is given by $-2\mu B^j_z\tau/\hbar$, where $\tau$ is the neutron's transit time in the magnetic field region, with $j=\alpha,\beta$.
The $\beta$ compensation and the spin analysis is realized only in the forward exit beam, normally corresponding to the $|+\rangle_{\chi=0}$ state. By changing the phase $\chi$ between the beam paths from 0 to $\pi$ we can however flip the meaning of the exit beams and thereby analyze also the $|-\rangle_{\chi=0}$ state. The spin analysis is realized by a magnetic super mirror which lets pass only the $|s_{z+}\rangle$ state. In combination with a $\pi /2$ spin rotator (DC2) it analyzes the $|s_{x+}\rangle$ state required here. The position of the DC2 coil was adjusted in beam direction to catch the precessing spin at the correct angle required for the $\pi/2$ rotation.

The experiment was carried out at the neutron interferometer instrument S18 at the high-flux reactor of the Institute Laue-Langevin (ILL) in Grenoble, France. A monochromatic beam with mean wavelength $\lambda=1.91$\AA ($\delta\lambda/\lambda\sim0.02$) and $5\,\times \,7\,\mathrm{mm^{2}}$ beam cross section was used. The experimental data is identified by \cite{S18data}.

\begin{figure}[!t]
\includegraphics[width=\columnwidth]{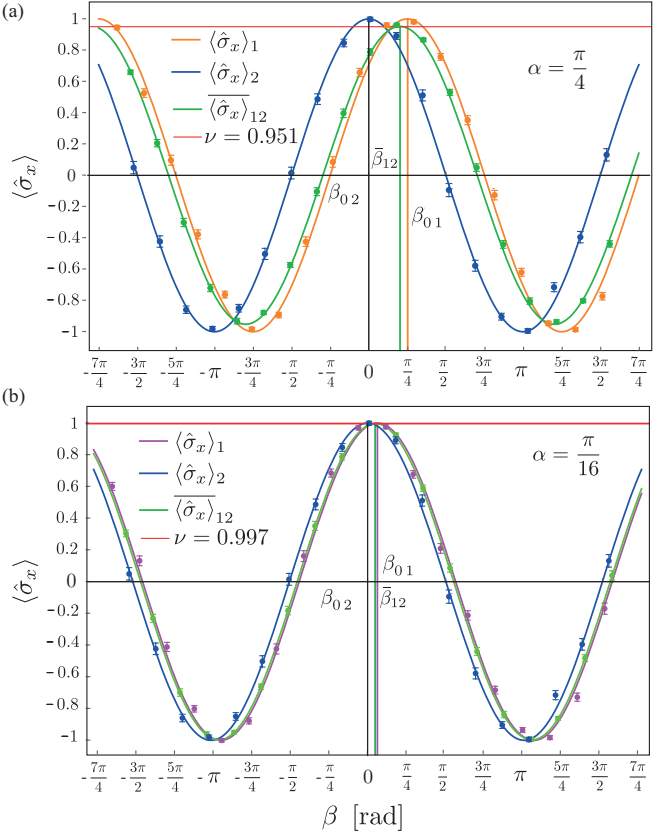}
	\caption{Experimental results of feedback compensation in which-way context. Measured phase shifts yield for (a) $\alpha=\pi/4:\,\beta_{0\, 1}^{\rm{meas}}=0.2533(61)\,\pi$  ($\beta_{0\, 2}^{\rm{meas}}=-0.0012(38)\,\pi$) and (b)  $\alpha=\pi/16:\,\beta_{0\, 1}^{\rm{meas}}=0.0646(66)\,\pi$. Solid curves and vertical lines indicate respective theoretical predictions and error bars represent $\pm 1$ standard deviation.}\label{fig:WWResults}
	\end{figure}

\subsection{Which-way context}
Formally the \emph{which-way} measurement is described by the observable $\hat\Pi_1$, i.e., the projector onto path 1. Experimentally the projector $\hat\Pi_{1,2}$ is realized by blocking path 2,1, which is schematically illustrated in Fig.\,\ref{fig:setup}. 
Depending on which projection is measured the respective compensation operation $\hat U^\beta_z=\textrm{exp}(-{\rm i}(\beta_{0\,1,2})/2\,\,\hat\sigma_z)$ is applied, with $\beta_{0\, 1}=-\alpha$ for path 1 and $\beta_{0\, 2}=0$ for path 2.
The obtained results of which-way measurements are plotted in Fig.\,\ref{fig:WWResults} (a) for $\alpha=\pi/4$ (strong interaction) and Fig.\,\ref{fig:WWResults} (b) for $\alpha=\pi/16$ (weak interaction). The experimentally observed phase shifts $\beta^{\rm{meas}}_{0\,1,2}$ (see caption Fig.\,\ref{fig:WWResults}) reproduce the theoretically predicted values from Eq. (\ref{eq:sxexpectweak}), given by 
\begin{eqnarray}
\langle\hat\sigma_x\rangle_1&=&\cos(\beta-\omega_{11}\,\alpha)\nonumber\\
\langle\hat\sigma_x\rangle_2&=&\cos(\beta-\omega_{12}\,\alpha) 
\end{eqnarray}
 (within the error-bars), accurately yielding $\omega_{11}=1$ and $\omega_{12}=0$. 
The average path presence is expressed as
 \begin{eqnarray}
\overline{\langle\hat\sigma_x\rangle}_{12}=p_1\cos(\beta-\omega_{11}\,\alpha)+p_2\cos(\beta-\omega_{12}\,\alpha),
\end{eqnarray}
with $p_1=0.8$ and $p_2=0.2$. The reduced visibility $\nu$ is indicated by red lines and amounts to $\nu\sim 1 -1/2\,\alpha^2 \mathit\Delta^2(\bar \omega_{1})$, with path uncertainty $ \mathit\Delta^2(\bar \omega_{1}) 
=p_1p_2=0.16$. 
The Experimentally obtained values of the path presence $\omega_1$ are summarized in Fig.\,\ref{fig:pp} for the interaction strengths $\alpha=\pi/4,\,\pi/8$ and $\alpha=\pi/16$.

\subsection{Interference context}

\begin{figure}[!b]
\includegraphics[width=\columnwidth]{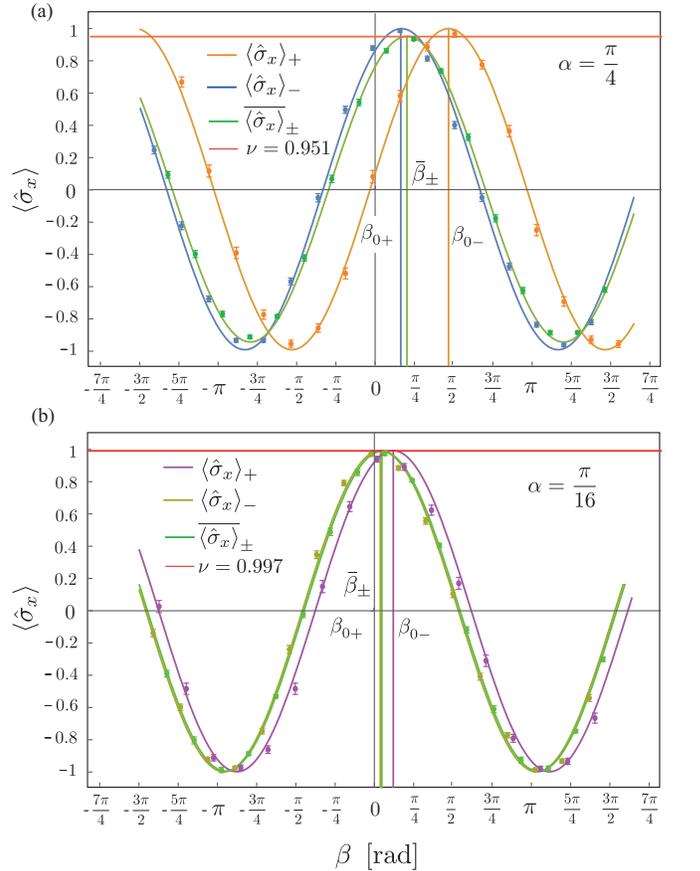}
	\caption{Experimental results of feedback compensation in interference context. Measured phase shifts yield (a) $\alpha=\pi/4:\,\beta^{\rm{meas}}_{0+}=0.1671(61)\,\pi$, $\beta^{\rm{meas}}_{0-}=0.4727(35)\,\pi$ and (b)  $\alpha=\pi/16:\,\beta^{\rm{meas}}_{0+}=0.0449(54)\,\pi$, $\beta^{\rm{meas}}_{0-}=0.1229(54)\,\pi$. Solid curves and vertical lines indicate respective theoretical predictions and error bars represent $\pm 1$ standard deviation.  }\label{fig:IntResults}
	\end{figure}

In the \emph{interference} context the optimal correction is given by $\beta_{0\pm}=-\omega_{1\pm}\alpha$ as explained in the previous Sections. Path presences of $w_{1+}=2/3$ and $w_{1-}=2$, with probabilities $p_+=0.9$ and $p_-=0.1$ respectively, are obtained in the limit of weak coupling (small $\alpha$).

%
%

Applying the respective setup from Fig.\,\ref{fig:setup}, the total combined state, consisting of object system (path) and probe system (spin), behind the phase shifter flag reads 
\begin{eqnarray}
\vert\Psi(\chi)\rangle_{\rm{tot}} &=&\cos\frac{2}{\sqrt 5}\vert 1\rangle\,{\textrm{exp}}(-{\rm i}\frac{\alpha}{2}\hat\sigma_z)\vert s_{x+}\rangle\nonumber\\&+&e^{\rm{i}\chi}\sin\frac{1}{\sqrt 5}\vert 2 \rangle \vert s_{x+}\rangle,
\end{eqnarray}
where $\chi$ is the relative phase between the (path) eigenstates, adjusted by a phase shifter. The phases $\chi=0$ and $\chi=\pi$ account for the post selection of state $|+\rangle_{\chi=0}$ and $|-\rangle_{\chi=0}$, respectively. Depending on the selected state, the respective compensation operation $\hat U^\beta_z=\textrm{exp}(-{\rm i}(\beta_{0\pm})/2\,\,\hat\sigma_z)$ is applied behind the interferometer.

The experimentally observed phase shifts $\beta^{\rm{meas}}_{0\pm}$ (see caption Fig.\,\ref{fig:IntResults}) evidently reproduce the feedback fringes from Eq. (\ref{eq:sxexpectweak}) expressed as
\begin{eqnarray}
\langle\hat\sigma_x\rangle_+&=& \cos(\beta-\omega_{1+}\,\alpha)\nonumber\\
\langle\hat\sigma_x\rangle_-&=& \cos(\beta-\omega_{1-}\,\alpha),
\end{eqnarray}
and the average is expressed as
\begin{eqnarray}
\overline{\langle\hat\sigma_x\rangle}_{\pm}= p_+\cos(\beta-\omega_{1+}\,\alpha)+p_-\cos(\beta-\omega_{1-}\,\alpha).
\end{eqnarray}

The obtained measurement results for measurement strength $\alpha=\pi/4$ are plotted in Fig.\,\ref{fig:IntResults}\,(a).
The corresponding output port probabilities are  
$p_+=0.8696$ and $p_-=0.1304$. 
The conditional compensation fringes are shifted by $\omega_{1\pm}\,\alpha$. For $\alpha=\pi/4$ the predicted values of the path presence are 
$\omega_{1+} = 0.6686$ at $p_+ = 0.8696$ and
$\omega_{1-}= 1.8701$ at $p_- = 0.1304$ 
and are evidently reproduced in the experiment, which can be seen from the values in the figure captions.
 The green fringes in the figures show the statistical average of all neutrons, averaged over both output channels. 

\begin{figure}[!t]
\includegraphics[width=\columnwidth]{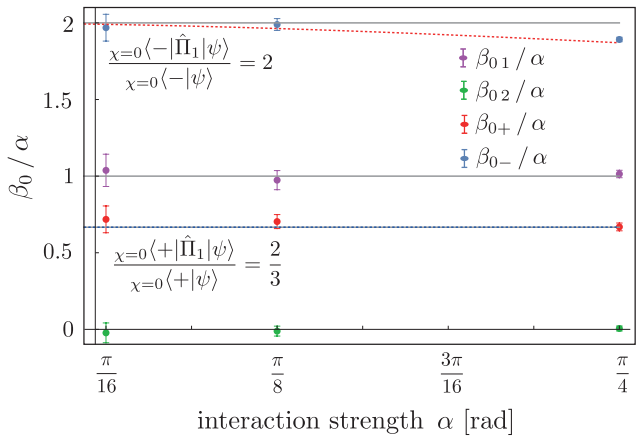}
	\caption{Optimal compensation $\beta_0/\alpha$ versus measurement strength $\alpha$ for $\alpha=\pi/4,\,\pi/8$ and $\pi/16$. Dotted lines represent exact theoretical predictions of $\omega_{1\pm}$ in interference context, deviating from weak values for larger values of $\alpha$. }\label{fig:pp}
	\end{figure}

A plot of the optimal compensation $\beta_0/\alpha$ versus interaction strength $\alpha$ is given in Fig.\,\ref{fig:pp}. The measured path presences converge for small $\alpha$ towards the ideal weak values. For large $\alpha$, the optimal compensation angle is no longer given by the first order term of Eq. (\ref{eq:beta0pmSeries}) but by the complete expression Eq. (\ref{eq:beta0pm}), shown by the dotted line in Fig. \ref{fig:pp}. In addition, a strong coupling between spin and path modifies the amplitudes in the exit beams, cf. Eq. (\ref{eq:ApmSeries}), and the probe qubit perturbs the actual measurement.

\section{Discussion} \label{sec:discussion}

Our experimental results show that spin rotations applied in one of the paths in a two path neutron interferometer result in proportional rotations of the spin in the output ports of the interferometer. Each of the two different rotation angles observed in the output ports are precisely defined and show no fluctuations. We therefore conclude that the spin rotations define a precise value of the presence of a particle in the path in which the rotation was applied. This result indicates that the particles have a precisely defined fractional presence in the path when interference effects are observed in the output ports. 

We would like to point out that this result is consistent with Ozawa's theory of measurement errors, which predicts an error of zero when the measurement results are given by the weak values defined by a pure input state and a precise projective measurement. The experiment confirms that the error measure introduced by Ozawa corresponds to experimentally observable fluctuations in the effects of weak interactions, strongly suggesting that the weak values are an accurate description of the dependence of physical properties on the measurement context \cite{Tollaksen07}. It seems to be highly significant that the feedback compensation approach establishes the consistency of Ozawa uncertainties, weak values and the quantum mechanics of weak unitary interactions. The experimental results show that weak values and Ozawa uncertainties provide an accurate description of the experimentally observable statistics of weak interactions.

Keen readers may ask where the spin rotations expressed in Eq.\,(22) actually come from. The $\alpha$ rotation in path 1 rotates the original $+x$ spin state slightly within the $xy$ plane. Expressing the spin by $x$ eigenstates, we obtain a small $-x$ component superimposed with the main $+x$ component. The amplitude of the $-x$ component carries the which-way information. When the components of the two paths interfere, the relation between the $+x$ and $-x$ components change, corresponding to a rotation of the Bloch vector of the spin. In general, this rotation depends on the interferometer phase $\chi$. For $\chi=0$ or $\chi=\pi$ the probe in the outgoing beams points along a direction in the $xy$ plane of the Bloch sphere.

In neutron interferometry we always have clear experimental evidence that the interference is based on self-interference of individual particles. Firstly, neutrons are fermions and would never occupy the same state, and secondly the beam intensity is so low that there are virtually never two particles simultaneously in the setup. Still, the visibility of interference itself requires a whole ensemble of neutrons, allowing alternative interpretations of quantum mechanics which assume locality of single particles. The present experiment rules out such statistical interpretations due to the vanishing error of the path presence determined by the weak values.

\section{Conclusion}

The precise analysis of the effects of sufficiently small spin rotations applied in only one path of a two path neutron interferometer shows that the presence of the neutrons in the paths during an interference experiment is accurately described by fractional values corresponding to the weak values associated with the output ports of the interferometer, where the precision of the results is given by the Ozawa uncertainties. 
We have experimentally verified that these fluctuations are close to zero in the present case, providing the first experimental evidence that the partial path presences described by weak values apply to each individual neutron detected in the corresponding output beam of the interferometer. Since this phenomenon is not observed when a which path measurement is performed, our results also demonstrate that the way in which particles propagate through an interferometer depends on the measurements performed in the output. The same initial uncertainty of the particle presence that appears as a statistical distribution of detection events in a which-path measurement will appear as the fluctuation of the partial presence of each individual particle when interference effects are detected instead.

It should be emphasized that all of these results are completely consistent with standard quantum theory. The conclusion that particles can be physically delocalized between paths in which no strong interactions occur and that the localization or delocalization is decided by a measurement that takes place after the particles have propagated along the paths is a possibility inherent in the paradoxical aspects of quantum superpositions. What is new in the present work is that we demonstrate that standard quantum theory predicts precise and specific effects of the presence of a particle in a path, even when the particle only undergoes a very weak interaction on its way though the interferometer.

\section{Acknowledgement}
 This work was supported by the Austrian science fund (FWF) Projects No. P 30677, P 34239, and P 34105.

\appendix*
\section{}
\subsection{Spin expectation values} \label{app:expect}

The spin expectation values in the directions $d \in \{x,y,z\}$ are given by
\begin{subequations}
\begin{alignat}{1}
\langle\hat\sigma_d\rangle &= 
  \frac{  (+1) \cdot |s_{d+}|^2 + (-1)\cdot |s_{d-}|^2 }
  {|s_{d+}|^2+|s_{d-}|^2}  \\
 s_{d\pm} &= \langle s_{d\pm} |  
\left( \omega_{1}\hat U_z^{\alpha-\beta} + \omega_{2}\hat U_z^{-\beta} \right)| s_{x+}\rangle
\end{alignat}
\end{subequations}
and with $\beta_{0\pm}$ and $A_\pm$ defined in Eqs. (\ref{eq:beta0pm}) and (\ref{eq:Apm}) respectively and
\begin{subequations}
\begin{alignat}{1}
   |s_{x+}|^2 &=\frac {|A_\pm|^2}2 \Bigl[\cos(\beta-\Re\beta_{0\pm})  
   + \cosh \Im \beta_{0\pm} \Bigr]\\
   |s_{x-}|^2 &=\frac {|A_\pm|^2}2 \Bigl[-\cos(\beta-\Re\beta_{0\pm})  
   + \cosh \Im \beta_{0\pm} \Bigr]\\
   |s_{y+}|^2 &=\frac {|A_\pm|^2}2\Bigl[\sin(\beta-\Re\beta_{0\pm})  
   + \cosh \Im \beta_{0\pm} \Bigr]\\
   |s_{y-}|^2 &=\frac {|A_\pm|^2}2\Bigl[-\sin(\beta-\Re\beta_{0\pm})  
   + \cosh \Im \beta_{0\pm} \Bigr]\\
   |s_{z+}|^2 &=\frac {|A_\pm|^2}2 \exp(\Im\beta_{0\pm})\\
   |s_{z-}|^2 &=\frac {|A_\pm|^2}2 \exp(-\Im\beta_{0\pm})
\end{alignat}
\end{subequations}
we obtain the results
\begin{subequations}
\begin{alignat}{1}
  \langle \hat \sigma_x \rangle &= 
  \frac {\cos (\Re\beta_{0\pm}-\beta)}{\cosh \Im \beta_{0\pm}}  \\
  \langle \hat \sigma_y \rangle &= 
  \frac {\sin (\Re\beta_{0\pm}-\beta)}{\cosh \Im \beta_{0\pm}}  \\
  \langle \hat \sigma_z \rangle &=
  \tanh \Im \beta_{0\pm}.
\end{alignat}
\end{subequations} 

\subsection{Variances of spin expectation values} \label{app:variances}

The variances $\mathit\Delta(\langle\hat\sigma_d\rangle)$ of the expectation values are given by 
\begin{subequations}
\begin{alignat}{1}
\mathit\Delta(\langle\hat\sigma_d\rangle) &= p_{d+} (+1 - \langle\hat\sigma_d\rangle)^2 + p_{d-} (-1 - \langle\hat\sigma_d\rangle)^2\\
p_{d\pm}&=  \frac{ |s_{d\pm}|^2 }
  {|s_{d+}|^2+|s_{d-}|^2}
\end{alignat}
\end{subequations}
and we obtain
\begin{subequations}
\begin{alignat}{1}
\mathit\Delta^2(\langle \hat \sigma_x \rangle) &= 
  \left( 1 - \frac {\cos^2 (\Re\beta_{0\pm}-\beta)}{\cosh \Im \beta_{0\pm}}\right)  \\
\mathit\Delta^2(\langle \hat \sigma_y \rangle) &= 
  \left( 1 - \frac {\sin^2 (\Re\beta_{0\pm}-\beta)}{\cosh \Im \beta_{0\pm}}\right)  \\
\mathit\Delta^2(\langle \hat \sigma_z \rangle) &=
  \frac 1 {\cosh \Im \beta_{0\pm}}.
\end{alignat}
\end{subequations} 
For optimal compensation ($\chi=0$ and $\beta=\Re\beta_{0\pm}$) we find that  $\mathit\Delta(\langle \hat \sigma_x \rangle)$ vanishes completely while $\mathit\Delta(\langle \hat \sigma_y \rangle)$ and $\mathit\Delta(\langle \hat \sigma_z \rangle)$ are maximal and equal unity.
For small $\alpha$ and $\beta=0$ this turns into
\begin{subequations}
\begin{alignat}{1}
\mathit\Delta^2(\langle \hat \sigma_x \rangle) &= 
  \alpha^2 |\omega_{1\pm}| \\
\mathit\Delta^2(\langle \hat \sigma_y \rangle) &= 
  1- \alpha^2 \Re\omega_{1\pm}^2 /2\\
\mathit\Delta^2(\langle \hat \sigma_z \rangle) &=
  1- \alpha^2 \Im\omega_{1\pm}^2 /2.
\end{alignat}
\end{subequations}

\subsection{Expectation values averaged over exit beams} \label{app:averaged}

In the interference context the expectation values $\langle\sigma_x\rangle$ in the two exit beams $|+\rangle$ and $|-\rangle$ are given by Eq. (\ref{eq:sigmax}). The average can be written as
\begin{subequations}
\begin{alignat}{1}
 \overline{\langle\hat\sigma_x\rangle}_{\pm} &= 
 p_+ \langle\hat\sigma_{x}\rangle_+  + p_-  \langle \hat\sigma_{x}\rangle_- = 
 \nu_{\pm} \cos(\beta - \bar\beta_{\pm}) \label{eq:sxexpectavrg}\\
  \bar\beta_{\pm} &= \arctan \frac{p_+ \sin\beta_{0-} + p_- \sin\beta_{0-}}{p_+ \cos\beta_{0-} + p_- \cos\beta_{0-}} \\
  \nu_{\pm}^2 &=1 - \frac{(p_1-p_2)^2}2 \left[1-\cos(\beta_{0+}-\beta_{0-})\right]
\end{alignat}
\end{subequations}
calculated for real weak values and $\chi=0$. A series expansion by $\alpha$ gives
\begin{subequations}
\begin{alignat}{2}
  \bar\beta_{\pm} &= p_1 \alpha + \mathcal{O}( \alpha^3)  \\
  \nu_{\pm} &= 1 - \frac 1 2 p_1 p_2 \alpha^2 +  \mathcal{O}( \alpha^4).
\end{alignat}
\end{subequations}

In the which-way context the expectation values in path 1 and 2 read $\langle\sigma_x\rangle_1=\cos(\alpha-\beta)$ and $\langle\sigma_x\rangle_2 = \cos(\beta)$ respectively, and the average is given by
\begin{subequations}
\begin{alignat}{1}
 \overline{\langle\hat\sigma_x\rangle}_{12} &=  p_1 \langle\hat\sigma_{x}\rangle_1  + p_2  \langle \hat\sigma_{x}\rangle_2
   = \nu_{12}  \cos(\beta - \bar\beta_{12}) \\
  \bar\beta_{12} &= \arctan \frac{p_1\sin\alpha}{p_2+p_1\cos\alpha} \\
  \nu_{12}^2 &=1- 2 p_1 p_2(1- \cos\alpha).
\end{alignat}
\end{subequations}
A series expansion by $\alpha$ gives
\begin{subequations}
\begin{alignat}{2}
  \bar\beta_{12} &= p_1 \alpha + \mathcal{O}( \alpha^3)
  \\
  \nu_{12} &= 1 - \frac 1 2 p_1 p_2 \alpha^2 +  \mathcal{O}( \alpha^4).
\end{alignat}
\end{subequations}

\pagebreak

\bibliography{feedbackcomp}

\end{document}